\documentclass[prl,a4paper,nofootinbib,
twocolumn,
]{revtex4}
\usepackage{graphicx}
\usepackage{amsfonts}
\usepackage{amssymb}
\usepackage{slashed} 
\usepackage{color}
\usepackage{hyperref}
\def\eq#1{{Eq.~(\ref{#1})}}
\def\eqs#1#2{{Eqs.~(\ref{#1})--(\ref{#2})}}

\def\Re{\mbox{Re}\,}

\def\di{\mbox{d}}
\def\ltap{\ \raisebox{-.4ex}{\rlap{$\sim$}} \raisebox{.4ex}{$<$}\ }

\definecolor{oucrimsonred}{rgb}{0.6, 0.0, 0.0}
\definecolor{persianblue}{rgb}{0.11, 0.22, 0.73}
\definecolor{forestgreen}{rgb}{0.13,0.35,0.13}
 \hypersetup{colorlinks, citecolor=oucrimsonred, linkcolor=persianblue, urlcolor=oucrimsonred}
\def\hhref#1{\href{http://arxiv.org/abs/#1}{#1}} % in bibliography
\newcommand{\be}{\begin{equation}}
\newcommand{\ee}{\end{equation}}
\newcommand{\bea}{\begin{eqnarray}}
\newcommand{\eea}{\end{eqnarray}}
\newcommand{\nn}{\nonumber}

\begin{document}
%%%%%%%%%%%%%%%%%%%%%%%%%%%%%%%%%%%%%%%%%%%%%%%%%%%%%%%%%%%  FRONT PAGE
\title[]{Hunting  down  massless dark photons in kaon physics
%$K^+\rightarrow \pi^+\pi^0 + \slashed{E}$
}
\date{\today}
\author{M.\ Fabbrichesi$^{\dag}$}
\author{E.\ Gabrielli$^{\ddag\dag}$}
\author{B.\ Mele$^{\ast}$}
%\author{XYZ}
\affiliation{$^{\dag}$INFN, Sezione di Trieste, Via  Valerio 2, 34127 Trieste, Italy }
\affiliation{$^{\ddag}$Physics Department, University of Trieste and  NICPB, R\"avala 10, Tallinn 10143, Estonia }
\affiliation{$^{\ast}$INFN, Sezione di Roma, P.le Aldo Moro 2, 00185 Roma, Italy}
\begin{abstract}
\noindent  If dark photons are massless, they couple to standard-model particles only via higher dimensional operators, while direct (renormalizable) 
%the kinetic mixing with photons
interactions induced by kinetic-mixing, which motivates most of the current experimental searches, 
%is
are absent.
We consider the effect of possible flavor-changing 
magnetic-dipole couplings of massless dark photons in kaon physics. In particular, we study the branching ratio for the process $K^+\rightarrow \pi^+\pi^0 \bar \gamma$   with a simplified-model approach, assuming the chiral quark model to evaluate the hadronic matrix element. 
Possible effects in the $K^0$-$\bar K^0$ mixing are taken  into account. 
We find that branching ratios up to $O(10^{-7})$ are allowed---depending  on the 
dark-sector masses and couplings.
Such large branching ratios for $K^+\rightarrow \pi^+\pi^0 \bar \gamma$ could be of interest for experiments dedicated to rare $K^+$ decays like NA62 at CERN, where $\bar \gamma$ can be detected as a massless invisible system.
\end{abstract}
%%%%%%%%%%%%%%%%%%%%%%%%%%%%%%%%%%%%%%%%%%%%%%%%%%%%%%%%%%%%%%%%%%%
\maketitle
%%%%%%%%%%%%%%%%%%%%%%%%%%%%%%%%%%%%%%%%%%%%%%%%%%%%%%%%%%%%%%%%%%%
The clarification of the origin of dark matter (DM) might require the existence of a  {\it dark sector}  made up of particles uncharged under the standard model (SM) gauge group. The possibility of extra secluded $U(1)$ gauge groups---mediating interactions in the dark sector via 
 {\it dark photons}--- 
is the subject of many experimental searches (see~\cite{Raggi:2015yfk} for recent reviews). These searches are mostly based on the assumption that the secluded $U(1)$ gauge group is broken, and 
the corresponding {\it massive} dark photon ($\gamma^\prime$) interacts directly with the SM charged fields through  renormalizable (dimension-four) operators induced by the kinetic mixing between dark and electromagnetic photons. Experimental results are then parametrized in terms of the dark-photon mass $m_{\gamma^\prime}$ and mixing parameter $\epsilon$,
with dark photon signatures that can either correspond to its decay into SM particles or assume
an invisible decay into extra dark fields. 
Because the induced operators have  dimension four, most studies necessarily explore regions where the couplings are very
%artificially 
small ({\it millicharges}). 

We 
address instead the case of an unbroken dark $U(1)$ gauge symmetry, with a {\it massless} dark photon ($\bar\gamma$). 
The role of massless dark photons in galaxy formation and dynamics has been  discussed  in~\cite{Foot:2004pa,Ackerman:mha,Fan:2013tia,Foot:2014uba,Heikinheimo:2015kra}. A strictly massless dark photon  is very appealing from the theoretical point of view. 
 Indeed, for massless dark photons it is possible~\cite{Holdom:1985ag} to define two fields, the dark and the ordinary photon, in such a manner that the dark photon only sees the dark sector. In this basis,
%the kinetic-mixing coupling 
% is not physical, and 
%can be rotated away  
ordinary photons  couple to both the SM and the dark sector--- the latter  with millicharged strength to prevent macroscopic effects.
Massless dark photons therefore interact with SM fields only through higher dimensional operators---typically suppressed by the mass scales related to new massive fields charged under the unbroken dark $U(1)$ gauge symmetry~\cite{Dobrescu:2004wz}---while their coupling constants can take natural values thanks to the built-in suppression associated to  the higher dimensional operators. 
This  makes 
the $\bar\gamma$ direct production   in SM particle scattering/decay small and unobservable, consequently evading most of the  search strategies for dark photons 
currently ongoing in laboratories. A possible exception is provided by the Higgs boson decay into dark photons in  the nondecoupling regime. 
This scenario has been considered in~\cite{Biswas:2016jsh}, where  observable $\bar\gamma$ production rates mediated by the Higgs decay $H\to \gamma \bar\gamma$ have been found  at the LHC in realistic frameworks~\cite{Gabrielli:2013jka,Gabrielli:2016vbb}.
Flavor-changing-neutral-current (FCNC) decays of heavy flavors into a massless dark photon, $f\to f^\prime \bar \gamma$, can offer other search channels with potentially observable rates~\cite{Dobrescu:2004wz,Gabrielli:2016cut}. 

Here we focus on FCNC effects induced by massless dark photons $\bar{\gamma}$ in  kaon physics,
 and discuss the change of picture with respect to the massive case.

%%%%%%%%%%%%%%%%%%%%%%%
\begin{table*}[ht!]
\small
\begin{center}
\vspace{0.2cm}
\begin{tabular}{|c|c|c|c|c|}
\hline
$Q_1, \, \tilde Q_1$ &$Q_2, \, \tilde Q_2$ &$Q_3, \, \tilde Q_3$ &$Q_4$ &$Q_5$ \cr
\hline
$ \bar d_L^\alpha \gamma_\mu s_L^\alpha\, \bar d_L^\beta \gamma_\mu s_L^\beta \, , \,  (L \leftrightarrow R)  $  &
$\bar d_R^\alpha s_L^\alpha\, \bar d_R^\beta s_L^\beta\, ,\, (L \leftrightarrow R)   $ &
$ \bar d_R^\alpha s_L^\beta\, \bar d_R^\beta s_L^\alpha\, ,\, (L \leftrightarrow R)  $ &
 $ \bar d_R^\alpha s_L^\alpha\, \bar d_L^\beta s_R^\beta $ & 
 $ \bar d_R^\alpha s_L^\beta\, \bar d_L^\beta s_R^\alpha   $    \cr
\hline
 $1/3 \; m_K f_K^2 B_1(\mu) $ &$  -5/2\; X_K m_K f_K^2 B_3(\mu)$ &$  1/24 \; X_K m_K f_K^2 B_3(\mu)$ & $1/4 \; X_K  m_K f_K^2  B_4(\mu) $& $1/12 \; X_K m_K f_K^2 B_5(\mu)$  \cr
\hline
$-1/24 \,C^2$  & $0$ & $ 1/12 \,C^2 $ & $1/6 \, C^2$  & $1/6\,  C^2$ \cr
\hline
\end{tabular}
\end{center}
\caption{\small In the first two rows, relevant operators  are numbered according to the notation in~\cite{Gabbiani:1996hi,Ciuchini:1998ix}. The matrix elements $\langle K^0 | Q_i | \bar K^0 \rangle$
 (in the vacuum insertion approximation for the renormalized operators $Q_i$ at the low energy scale $\mu=2$ GeV)  are given in the third row multiplied by the respective bag factors $B_i(\mu)$~\cite{Ciuchini:1998ix} evaluated at same scale, with $X_K(\mu)=\left( m_K/(m_d(\mu)+m_s(\mu)) \right)^2$. The fourth row gives the Wilson coefficients at the matching scale (the common factor at the matching being  $C^2 = \xi^2/(16 \pi^2 \Lambda^2)$~\cite{Biswas:2016jsh}, where $\xi=g_L g_R/2$). Following \cite{Ciuchini:1998ix}, we take $m_d(\mu)=7$ MeV, $m_s(\mu)=125$ MeV, $m_K=497$ MeV, $f_K=160$ MeV, and $B_{1,2,3,4,5}(\mu)=0.60,0.66,1.05,1.03,0.73$, respectively. }
\label{tab:ops}
\end{table*}
%%%%%%%%%%%%%%%%%%%%%%%

The kaon system can be studied with great accuracy, allowing us to probe indirectly energy scales as large as tens of TeV, hence crucially constraining possible SM extensions. 
The   detection of massive dark photons in $K$ decays is presently under 
scrutiny~\cite{Raggi:2015yfk,Pospelov:2017pbt}. %Assuming kinetic mixing,
One can consider {\it radiative} $K$ decays where the (off-shell) SM photon $\gamma$ is replaced by a $\gamma^\prime$, and look  for 
%involving dark photons (or lepton pairs) $K\to X+ \gamma^\prime\;(\ell^+\ell^-)$ in place of the corresponding SM
%$K\to X+ \gamma\;(\ell^+\ell^-)$  channels, and look  
 resonances at $m_{\gamma^\prime}$ for either  $e^+e^-$ ($\mu^+\mu^-$) final states, or (in case $\gamma^\prime$ decays  into dark particles) for invisible final systems with a peak structure at $m_{\gamma^\prime}$ in the missing mass distribution.  
%both matching $\gamma^\prime$ masses  allowed by the $K$ phase space. 
Particular  emphasis has been given to  the decays $K^+\to \pi^+\gamma^\prime$ and $K^+\to \mu^+ \nu \, \gamma^\prime$~\cite{Pospelov:2008zw,Barger:2011mt,Carlson:2013mya,Batley:2015lha,Chiang:2016cyf}. However,
if  the secluded $U(1)$ gauge group is unbroken, 
 these two channels are not viable. Indeed, $K^+\to \pi^+\bar\gamma$ violates angular momentum conservation, while 
$K^+\to \mu^+ \nu \, \bar\gamma$ would require   unsuppressed  $\bar\gamma$ couplings. 

Because  $K^+\!$ decays into  a dark photon $\bar\gamma$ 
%plus anything  
must necessarily proceed through short-distance effects,
we  argue that  the most interesting channel to look for massless dark photons in kaon physics could be the decay $K^+\to \pi^+\pi^0\bar\gamma$. This decay  can be mediated by the FCNC transition $s\to d \, \bar \gamma$, prompted by a magnetic-dipole-type coupling
generated at one loop by the dark-sector degrees of freedom.
The dark photon gives rise in this case to a 
massless missing-momentum system inside the final state.
Recently, the sensitivity of the NA62 experiment at the CERN SPS~\cite{NA62:2017rwk} to two-body $K$ decays into a light vector decaying invisibly [$K^+\!\to \pi^+ \!+ \!(\gamma^\prime \!\to \!E_{miss})$] has been emphasized~\cite{Pospelov:2017pbt}. For the three-body $K^+\to \pi^+\pi^0\bar\gamma$ channel, whose kinematics is less characterized, the detection efficiency is expected to be  less favorable.  Nevertheless---since the $K^+\to \pi^+\pi^0\bar\gamma$ channel has a unique potential to unveil 
the existence of a massless dark photon---we think that the NA62 Collaboration should consider search strategies aiming at detecting this newly proposed  process, whose 
branching ratio (BR)  can reach $10^{-7}$ in  a {\it simplified} model of the dark sector, 
as  we estimate in the following.
%%%%%%%%%%%%%%%%%%%%

%%%%%%%%%%%%%%%%%%%%%%
\vskip0.3em
{\it A simplified model of the dark sector.}---We estimate  BR($K^+\to \pi^+\pi^0\bar\gamma$) in a simplified model that makes as few assumptions as possible, while providing the dipole-type transition we are interested in. 

The minimal choice in terms of fields consists of a SM extension  where there is a new (heavy) dark  fermion $Q$, singlet under the SM gauge interactions, but charged under an unbroken $U(1)_D$ gauge group associated to the massless dark photon. 
%Being massless, the corresponding dark photon does not have any renormalizable (tree-level) coupling to the SM fields \cite{Holdom:1985ag}.
SM fermions couple to  the dark fermion by means of a Yukawa-like interaction in the Lagrangian ${\cal L}$
\be
{\cal L} \, \, \supset \, \,  g_L (\bar{Q}_L q_R) S_R + g_R (\bar{Q}_R q_L) S_L + H.c. \; ,\label{lag}
\ee
where new (heavy) {\it messenger} scalar particles, $S_L$ and $S_R$, enter as well.  In \eq{lag},  the $q_L$ and $q_R$ fields are the SM fermions [$SU(3)$ triplets   and, respectively, $SU(2)$ doublets and singlets]. Flavor indices are  implicit, and we assume common ({\it i.e.\,}flavor blind) couplings $g_L$ and $g_R$.
 The left-handed messenger  field $S_L$  is a $SU(2)$ doublet, the right-handed messenger field  $S_R$ is a $SU(2)$ singlet, and both are $SU(3)$ color triplets.  These messenger fields are charged under $U(1)_D$, carrying the same $U(1)_D$ charge of the dark fermion.

In order to generate chirality-changing processes we also need in the Lagrangian the mixing terms
 
\be
{\cal L} \, \, \supset \, \, \lambda_S S_0 \left(  S_L  S_R^{\dag} \tilde H^\dag +  S_L^{\dag} S_R H \right) \, , \label{mix} 
\ee
where $H$ is the SM Higgs boson, $\tilde{H}=i\sigma_2 H^\star$, and $S_0$ a scalar singlet. The Lagrangian in \eq{mix} gives rise to  the mixing after both  the $S_0$ and $H$ scalars  take a vacuum expectation value (VEV), respectively, $\mu_S$  and $v$---the electroweak VEV. 
%The interaction Lagrangian in \eqs{lag}{mix} provides a simplified model  of the dark sector  and the minimal framework  to generate the
%$s\to d \; \bar{\gamma}$ transition.
%This model shares many similarities with the class of models describing the dynamical generation of exponentialy spread Yukawa couplings
%\cite{Gabrielli:2013jka,Gabrielli:2016vbb}, where a similar structure for the Lagrangian dark sector was required, except that two messengers doublets and singlets for each quark generation were introduced.
After diagonalization, the messenger fields $S_\pm$ couple  to both left- and right-handed SM fermions with strength $g_L/\sqrt{2}$  and $g_R/\sqrt{2}$, respectively. We can assume that the size of this mixing---proportional to the product  of the VEVs  ($\mu_s v$) ---is large and of the same order of the masses of the heavy fermion and scalars. 

The SM Lagrangian plus the  terms in \eqs{lag}{mix}  (supplemented by the corresponding kinetic terms) provide a simplified model for the dark sector and the effective interaction of the SM degrees of freedom with the massless dark photon $\bar\gamma$. 
SM fermions couple to $\bar\gamma$ only via nonrenormalizable interactions, induced by loops of the dark-sector  states.
Two scales are relevant: the dark fermion mass $M_Q$, which parametrizes the chiral symmetry breaking in the dark sector, and the lightest-messenger mass scale $m_S$. Since we are considering the contribution to the magnetic dipole operator (assuming vanishing 
quark masses), the dominant effective scale associated with it will either be chirally suppressed (being proportional to $M_Q/m_S^2$, for $m_S \gg M_Q$), or scale as $1/M_Q$ (for $m_S \ll M_Q$) due to decoupling. 
%Therefore, 
In order to have only one dimensionful parameter, in our analysis we assume a common mass for  the  dark fermion and the lightest scalar field, which we identify with the 
new-physics scale $\Lambda$. This choice   corresponds  to the maximum chiral enhancement.

This scenario is a simplified version of the model in \cite{Gabrielli:2013jka, Gabrielli:2016vbb, Gabrielli:2016cut} (possibly providing a natural solution to the SM flavor-hierarchy problem),  as well as a template for many models of the dark sector.

\vskip0.3em
{\it Bounds from $K^0$-$\bar K^0$ and astrophysics.}---A most  stringent limit to the mass scale and couplings of the above simplified 
model comes from its extra contributions to the $K^0$-$\bar K^0$ mixing in the kaon system   (related to the mass difference $\Delta M_{K}$ of  the neutral mass eigenstates $K_L$ and $K_S$, assuming $CPT$). 
%These must be consistent with $\Delta M_{K}  =  (3.484 \pm 0.006)\times 10^{-12}   \mbox{MeV}$~\cite{PDG}.

In order to compute the  dark-sector effects on $\Delta M_{K}$, we need to evaluate the dark-sector contribution to the effective Hamiltonian for the $\Delta S=2$
transitions, ${\cal H}^{\Delta S=2}_{eff}$
\be
\Delta M_{K}  = 2 \Re [ \langle K^0 |{\cal H}^{\Delta S=2}_{eff}
  | \bar K^0 \rangle]   \, .
\label{deltaMK}
\ee
The scalar-fermion interaction in \eq{lag} induces a new set of operators, which are reported in Table~\ref{tab:ops}, then obtaining 
\be
{\cal H}^{\Delta S=2}_{eff} = \sum_{i}^5  C_i Q_i + \sum_{i=1}^3 \tilde C_i \tilde Q_i   \, .
  \ee
  The  Wilson coefficients at the  matching scale  are computed by considering the exchange of the lightest messenger state in the loop, which provides a good estimate of the dominant contribution in the large-mixing limit of the
messenger mass sector.

We  compute the corresponding Wilson coefficients $C_i(\mu)$ at the ${\cal O}(\alpha_s)$ next-to-leading order,  after running them from the matching scale down to the low energy scale $\mu \sim 2$ GeV, where  the corresponding matrix elements are estimated on the lattice~\cite{Ciuchini:1998ix}.  We assume as matching scale the characteristic mass $\Lambda$ of the lightest-messenger and dark-fermion states, assumed  to be equal. 
Following this procedure, the dark-sector contribution to $\Delta M_{K}$ (in TeV) is
\be
 \Delta M_{K} = 8.47 \times 10^{-13} \frac{\xi^2}{\Lambda^2}\, , \label{con}
 \ee
 where  $\xi = g_L g_R/2$, and $\Lambda$ is  in TeV units. We  then assume that the above contribution of the new operators to \eq{deltaMK}    does not exceed 30\% of  the measured $\Delta M_{K}$ value~\cite{PDG}. Eq.~(\ref{con}) turns then into an upper bound for the allowed values for the $\xi^2/\Lambda^2$ ratio.
 %This gives an upper limit on the combination $\xi^2/\Lambda^2$.

While the flavor-changing dipole operator induced in the simplified model (see \eq{Hmagnetic} below) 
\textit{per se} is only bounded by kaon physics, if we make the (very conservative) assumption that the model also gives flavor-diagonal dipole operators and these are the same  size in the quark and lepton sectors, a bound can be derived from stellar cooling carried out by the emission of massless dark photons. Under these assumptions, the limit from $K^0$-$\bar K^0$ mixing in \eq{con} falls between the current astrophysical bounds~\cite{Hoffmann:1987et}---with the most stringent one from  white dwarves  being 1 order of magnitude stronger and that from the Sun 1 order of magnitude weaker.

\begin{figure}[ht!]
\begin{center}
\includegraphics[width=2.7in]{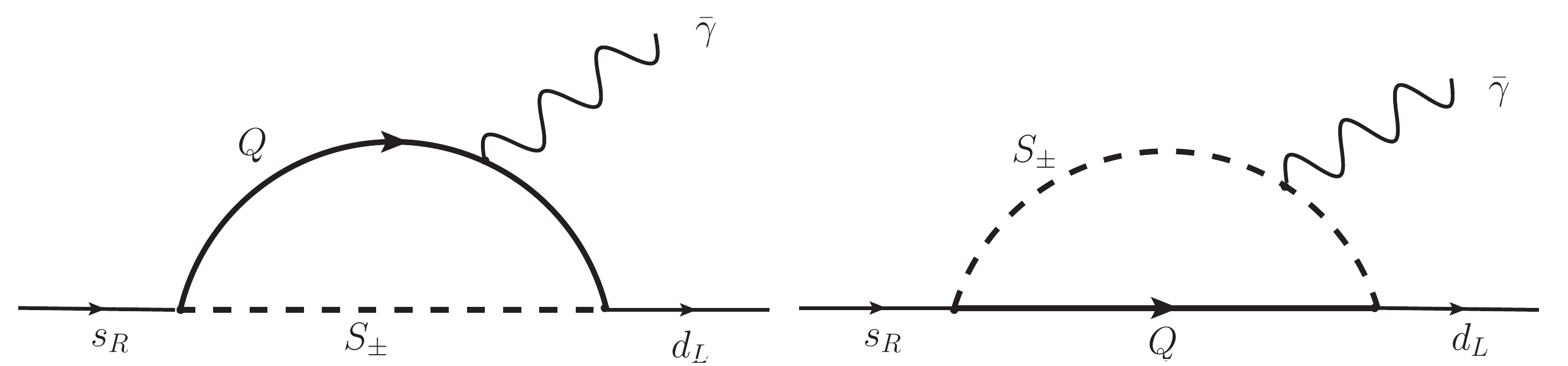}
 \caption{\small Vertex diagrams for the generation of the dipole operator in the simplified model of the dark sector (same for the specific model in \cite{Gabrielli:2013jka,Gabrielli:2016vbb,Gabrielli:2016cut}). 
\label{vertex} }
\end{center} 
\end{figure}

\vskip0.3em
{\it Amplitude and decay rate.}---The $K^+ \rightarrow \pi^+  \pi^0 \bar{\gamma}$  decay originates from the dimension-five magnetic dipole operator $\hat{Q}= \left(\bar s \, \sigma^{\mu\nu} \, d\right)   \bar{F}_{\mu \nu}$, where $\bar{F}_{\mu \nu}$ is the $\bar{\gamma}$ field strength, $\sigma_{\mu \nu}=\frac{1}{2}[\gamma_{\mu},\gamma_{\nu}]$,  and color and spin contractions are understood. $\hat{Q}$ enters  the effective Hamiltonian for $\Delta S=1$ transitions as
\be
   {\cal H}_{eff}^{\Delta S=1} =  \frac{e_D }{64 \,  \pi^2} \,  \frac{ \xi}{\Lambda} \,\hat Q\, ,
     \label{Hmagnetic}
\ee
where   $\alpha_D=e_D^2/(4\pi )$ is the $\bar{\gamma}$ coupling strength. The Wilson coefficient multiplying the magnetic operator in \eq{Hmagnetic} is obtained by  integrating the vertex function in our simplified model (see Fig.~\ref{vertex}). We have checked \eq{Hmagnetic}  by means of \texttt{Package X}~\cite{Patel:2015tea}.

%%%%%%%%%%%%%%%%%%%
%%%%%%%%%%%%%%%%%%%%%% 

The operator in \eq{Hmagnetic}  contributes only to the magnetic component of the process
\be
K^+ (p) \rightarrow \pi^+(q_1) \, \pi^0 (q_2)  \, \bar{\gamma} (k)\, , 
\ee
while its contribution to  the process $K^+\rightarrow \pi^+ \bar{\gamma}$  identically vanishes. The amplitude  \mbox{$\hat M \equiv \langle   \bar \gamma \; \pi^+  \pi^0 |\, {\cal H}_{eff}^{\Delta S=1}  \, | K^+ \rangle $} in the momentum space can be written as  
\be
\hat M =\frac{ M(z_1,z_2)}{m_K^3} \varepsilon_{\mu\nu\rho\sigma} q_1^\nu q_2^\rho k^\sigma \varepsilon^{\mu}(k)\, , \label{matrix}
\ee
where 
$\varepsilon^{\mu}(k)$ is the $\bar{\gamma}$ polarization vector. The corresponding    differential decay rate is
\bea
\frac{\di^2\Gamma}{\di z_1 \di z_2} &=& \frac{m_K}{(4\pi)^3} \left| M(z_1,z_2)\right|^2  \left\{ z_1 z_2 \left[1 - 2 (z_1+z_2) \right. \right. \nn \\ 
&  &\;\;\;  - \left.  \left. r_1^2 - r_2^2 \right] -r_1^2z_2^2 - r_2^2 z_1^2 \right\} \, , \label{gamma}
\eea
where $z_i=k\cdot q_i/m_K^2$ and $r_i = M_{\pi_i}/m_K$~\cite{Cirigliano:2011ny}. 

%%%%%%%%%%%%%%%%%%%
\begin{figure}[ht!]
\begin{center}
\includegraphics[width=2.7in]{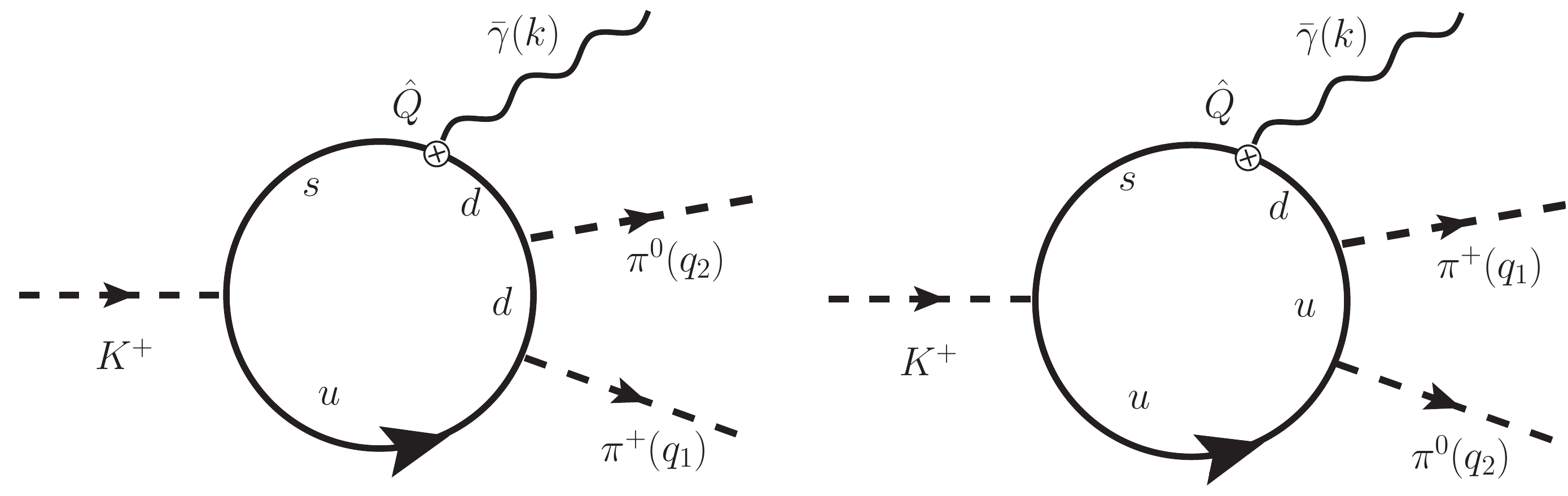}
\caption{\small $\chi$QM diagrams for the process $K^+  \rightarrow \pi^+ \pi^0   \tilde \gamma$.  The crossed circle stands for the insertion of the magnetic dipole operator $\hat Q$ in \eq{Hmagnetic}.
\label{chiQM} }
\end{center}
\end{figure}
%%%%%%%%%%%%%%%%%%%%%% 

  The matrix element in \eq{matrix}
%$\langle   \bar \gamma \; \pi^+  \pi^0 |\,  \hat Q \, | K^+ \rangle$
 can be estimated by means of the {\it chiral quark model} ($\chi$QM)~\cite{chiQM}.  In this model quarks are coupled to hadrons by  an effective interaction so that matrix elements can be evaluated by loop diagrams (see Fig.~\ref{chiQM}).
 In general there are  several free parameters, but in the present case only $M$, the mass of the constituent quarks, and $f$, the pion decay constant, enter the computation. The model has been applied to kaon physics in \cite{Antonelli:1995nv}, where a fit of the $CP$ preserving amplitudes of the nonleptonic decay of  neutral kaons has yielded a value $M = 200$ MeV~\cite{Bertolini:2000dy} with an error of less of 5\%.

According to the $\chi$QM we obtain that the magnetic component generated by the dipole operator in \eq{Hmagnetic} is given by
 \begin{widetext}
\bea
\frac{M(z_1,z_2)}{m_K^3} & =&    \frac{ e_D}{32\, \pi^2} \, \frac{\xi}{\Lambda} \,  \frac{M^3}{\pi^2 f^3 }    
 \; \Big[ M^2 D_0 (0, m_\pi^2 , m_\pi^2, m_K^2; 2 m_K^2 z_1 + m_\pi^2, m_K^2 (1-2 z_1 - 2 z_2) ; M, M, M, M)  \Big. \nn \\
&- &\Big.   D_{00}(0, m_\pi^2 , m_\pi^2, m_K^2; 2 m_K^2 z_1 + m_\pi^2, m_K^2 (1- 2 z_1 - 2 z_2 ) ; M, M, M, M) + (z_1 \leftrightarrow z_2)\Big]  \, .  \label{mm}
\eea
\end{widetext}
where $D_0$ and $D_{00}$ are four-point Passarino-Veltman coefficient functions  (see \cite{Ellis:2011cr} for their explicit form) to be evaluated numerically~\cite{Patel:2015tea}.

Inserting the amplitude in \eq{mm} in the differential decay rate in  \eq{gamma} yields, after integration and by normalizing $\Gamma$ by the total $K^+$ width  $\Gamma_{\rm tot}= 5.317 \times 10^{-14}$ MeV~\cite{PDG}, 
\be
{\rm BR}(K^+  \rightarrow \pi^+ \pi^0    \bar{\gamma} ) \simeq 1.31\;  \alpha_D\; \eta^2 \,\frac{\xi^2}{\Lambda^2} \, ,
\ee
where we assumed   $M=200$, $f=92.4$, $m_K=494$, and $m_{\pi^+} = m_{\pi^0} = 136$ MeV. 
 The coefficient $\eta$ accounts for the renormalization  of the Wilson coefficient of the dipole operator in going from the $\Lambda$ scale  to approximately $m_K$. We assume it equal to~1, and discuss the impact of possible uncertainties below.
 
\begin{figure}[ht]
\begin{center}
\includegraphics[width=3in]{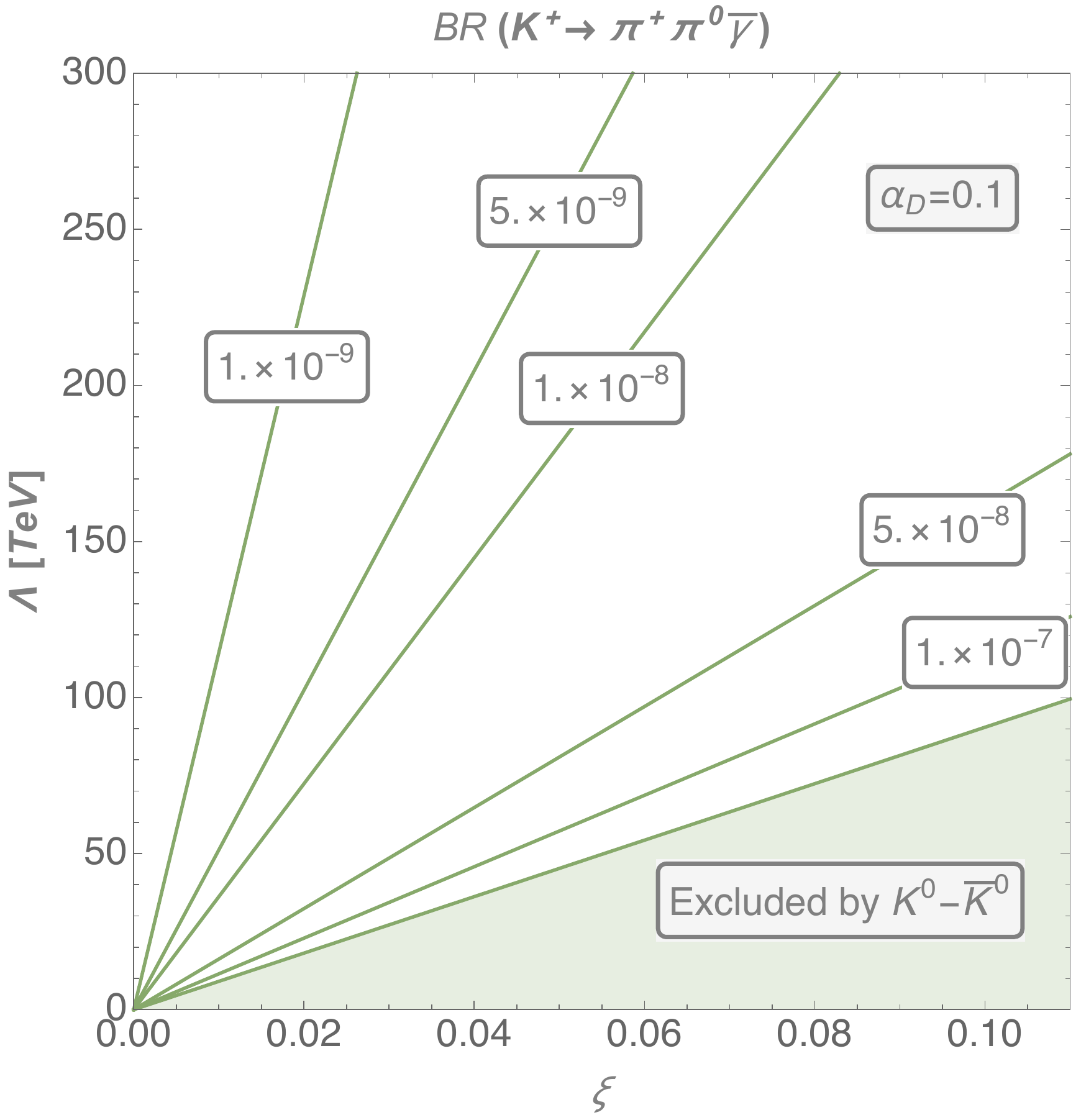}
\caption{\small BR($K^+\rightarrow \pi^+\pi^0 \bar{\gamma}$) as a function of the effective scale $\Lambda$ and coupling $\xi=g_L g_R/2$, for a  representative choice of the coupling strength $\alpha_D=0.1$. 
%Values  in the shaded region are excluded by the $K^0$-$\bar K^0$ constrain.
\label{BR} }
\end{center}
\end{figure}

  BR$(K^+  \rightarrow \pi^+ \pi^0    \bar{\gamma} )$ is proportional to $\xi^2/\Lambda^2$, 
  just as $\Delta M_{K}$ in Eq.~(\ref{con}). 
By taking for $\xi^2/\Lambda^2$  the value that saturates the $\Delta M_{K}$ constraint, we find an upper bound for  the BR which is, for the representative value \mbox{$\alpha_D=0.1$}, 
\be
{\rm BR}(K^+  \rightarrow \pi^+ \pi^0    \bar{\gamma} ) \ltap   1.6 \times 10^{-7} \, . \label{ub}
\label{boundBR}
\ee
 Fig.~\ref{BR} shows the BR$(K^+  \rightarrow \pi^+ \pi^0    \bar{\gamma} )$ contour plot versus  the scale $\Lambda$ and the coupling  $\xi$, for  $\alpha_D=0.1$. We see that a rather large range of parameters is allowed for which the BR is  sizable. The upper bound---given by  \eq{ub}---is represented in  Fig.~\ref{BR} by the boundary of the gray area.

There are three main sources of uncertainties in the result in \eq{ub}:
\begin{itemize}
\item The  matrix element estimate computed in the $\chi$QM depends on the parameter $M$. The result in \cite{Bertolini:2000dy} seems to indicate a rather small uncertainty on this parameter but one must be aware of the dependence. We find an increase by a factor  2.5 in the BR when going from $M= 200$ to 250 MeV; 
\item Even though there are $O(p^4)$ chiral perturbation theory  corrections to  $K^+\rightarrow \pi^+\pi^0 \bar{\gamma}$, these have been shown to be small~\cite{Ecker:1991bf}; 
\item By taking the QCD leading-order multiplicative value $\eta = 0.5$ (at $\mu=$ 2 GeV)~\cite{Buras:1993xp}, we find a   BR smaller by a factor 1/4. However, it is known that nonmultiplicative corrections go the opposite direction, and we thus need the   (not yet available) complete evolution before trusting this correction. Moreover, the QCD renormalization   introduces a strong dependence on the low-energy  scale $\mu$,  because the matrix element computed within the $\chi$QM is scale independent.
\end{itemize}
On top of these uncertainties, we have the overall dependence on the $\alpha_D$ strength  on which  the BR depends linearly. There exist cosmological relic density bounds on the ratio $\alpha_D/\Lambda^2$~\cite{Ackerman:mha}. Our choice of $\alpha_D=0.1$ is then consistent with $\Lambda$ of the order of 10 TeV.

%%%%%%%%%%%%%%%%%%%
%%%%%%%%%%%%%%%%%%%%%%

Similar predictions 
%for the $BR(K^+\rightarrow \pi^+\pi^0 \bar{\gamma})$ 
can be  obtained in the specific flavor model of \cite{Gabrielli:2013jka, Gabrielli:2016vbb, Gabrielli:2016cut}. In particular, for  $\alpha_D=0.1$, the approximate upper bound is given by
BR $\simeq 1.2 \times 10^{-8}$.
%, for $\xi=10^{-2}$ and $\Lambda \simeq 1 $ TeV. 
The  lower BR  is explained by the dark-fermion masses being related  in this case to the radiative generation of SM Yukawa couplings, resulting in a stronger chiral suppression of the effective scale associated with the dipole operator $\hat{Q}$, which turns out to be proportional to the bottom-quark  Yukawa coupling~\cite{Gabrielli:2016cut}.

\vskip0.3em 
{\it Conclusions.}--- NA62  at the CERN SPS will soon provide  a sample of 10$^{13}$ $K^+$,
with hermetic photon coverage and  good missing-mass resolution~\cite{NA62:2017rwk}.
We propose to look for the rare decay $K^+\rightarrow \pi^+\pi^0 \bar{\gamma}$ (where $\bar{\gamma}$ gives rise to a massless invisible system) as a sensitive probe for massless dark photons, for which the presently most explored dark-photon channels mediated by kinetic-mixing interactions in kaon decays are nonviable.

%\vskip 0.5em
%%%%%%%%%%%%%%%%%%%%%%%
%
%\begin{acknowledgments}
%EG would like to thank the CERN Theoretical Physics Department for its kind
%hospitality during the preparation of this work. MF is associated to SISSA and the Department of Physics, University of Trieste.  
%\end{acknowledgments}

%%%%%%%%%%%%%%%%%%%%%%%%%%%%%%%%%%%%%%

%%%%%%%%%%%%%%%%%%%%%%%%%%%%%%%%%%%%%%%%%%%%%%

\end{document}